\documentclass[prc,tightenlines,a4paper,nofootinbib]{revtex4}

\usepackage{amsmath,amssymb,graphicx}

\begin{document}

\title{Optical potentials, retarded Green's functions and nonorthogonality}

\author{Michael C. Birse}

\affiliation{Theoretical Physics Division, Department of Physics and Astronomy,
The University of Manchester, Manchester, M13 9PL, UK}

\begin{abstract}

An approach is outlined to constructing an optical potential that includes the effects of antisymmetry and target recoil. it is based on the retarded Green's function, which could make it a better starting point for applications to direct nuclear reactions, particularly when extended to coupled channels. Its form retains a simple connection to folding potentials, even in the presence of three-body forces.

\end{abstract}

\maketitle

\section{Introduction}

Formal treatments of the nuclear optical potential typically start from a Green's function for a projectile nucleon interacting with a nucleus. Different versions can be constructed starting from different choices of Green's functions:  time-ordered (or ``causal" or ``particle-hole"), retarded (or ``particle"), advanced (or ``hole").
The relationships between them have been reviewed at length by Capuzzi and Mahaux \cite{Capuzzi:1995}, and also by Escher and Jennings \cite{Escher:2002ud}. 
 Cederbaum gives further discussion including an extension that is relevant to approach developed here \cite{Cederbaum:2001}. At higher energies, potentials based on multiple-scattering theory \cite{Kerman:1959fr} can be used, as in Ref.~\cite{Vorabbi:2015nra}.
Surveys of modern microscopic approaches to optical potentials can be found in  recent reviews by Dickhoff and Charity \cite{Dickhoff:2018wdd} and Rotureau
\cite{Rotureau:2020ncy}. A broader overview of microscopic approaches to nuclear scattering and reactions can be found in the FRIB Theory AllIance white paper \cite{Johnson:2019sps}, which also examines their relationships to many-body methods used in calculations of nuclear structure.

The most commonly used approach starts from the time-ordered Green's function \cite{Bell:1996mz} (for a textbook treatment, see Ch.~14 of Ref.~\cite{Blaizot:1985}). 
This is based on many-body treatments of infinite matter and it can work well for heavier nuclei where recoil effects may be neglected. Recent applications to microscopic optical potentials to low-energy scattering can be found in Refs.~\cite{Rotureau:2016jpf,Whitehead:2018bfs,Idini:2019hkq}. These start with the same chiral two- and three-nucleon forces \cite{Epelbaum:2008ga} that are now used in calculations of nuclear structure. 

In contrast, approaches based on scattering theory start from the retarded Green's function, or the corresponding scattering wave function \cite{Feshbach:1958nx,Feshbach:1962ut,Villars:1967,Redish:1970}. These have more in common with the few-body techniques used to describe nuclear reactions.  
For example, they can handle the effects of recoil, which can be important for scattering and reactions on lighter nuclei. Although this was explored in an early paper by Redish and Villars \cite{Redish:1970}, it has only recently been revisited by Johnson \cite{Johnson:2017xdg,Johnson:2019qbe}. In the present work, I develop a version of this approach that describes coupled channels and so could be applied to direct reactions. Unlike other versions, the resulting optical potential includes pieces with the forms of folding potentials obtained from both two- and three-nucleon forces.

One complication for any treatment is the indistinguishability of projectile and target nucleons which means that the natural basis states to describe scattering are not orthogonal. This is handled rather differently in the various frameworks
\cite{Capuzzi:1995,Escher:2002ud,Cederbaum:2001}. Johnson's work \cite{Johnson:2017xdg,Johnson:2019qbe} examined the role of antisymmetry of the projectile-plus-target wave function as well as showing how translational invariance could be restored. In addition, he proposed an extension of these ideas to a coupled-channel treatment \cite{Johnson:2019qbe} which is closely related to the one studied by Cederbaum \cite{Cederbaum:2001}. These approaches start from a matrix of off-diagonal or inelastic Green's functions for a single nucleon in the presence of an $A$-body target nucleus, considering all initial and final states of that nucleus \cite{Cederbaum:2001,Johnson:2019qbe}. As a result, they have the potential to describe reactions as well as elastic scattering.

The off-diagonal Green's functions satisfy a set of coupled-channel equations with a matrix generalisation of the optical potential. The Born term in that potential differs from the usual ``folding" term by a numerical factor. Moreover, the corresponding numerical factors are different for three- and more-body forces and so each type of force has to be treated individually. This means that there is no simple connection to the usual folding term, especially if the microscopic Hamiltonian includes more that two-body forces. This is a concern because three-body forces are essential elements of chiral effective field theories \cite{Epelbaum:2008ga} and many-body forces are also generated when the similarity renormalisation group is used to ``soften" interactions, as is now frequently done in calculations of nuclear structure~\cite{Hergert:2015awm}.

The starting point for the present approach is a set of off-diagonal retarded Green's functions that is similar to those in Refs.~\cite{Cederbaum:2001,Johnson:2019qbe}. This works with the complete set of states consisting of a projectile nucleon with definite momentum and an $A$-body nucleus in one of its energy eigenstates. By inserting this complete set of these states into the equation for the off-diagonal Green's functions, I obtain a set of coupled equations in an effective one-body space. This space is spanned by the channels of relative motion of the nucleon and the target. 

Feshbach's projection formalism \cite{Feshbach:1958nx,Feshbach:1962ut} can then be used to eliminate all but a small number of these channels, giving an optical potential (or its matrix generalisation). Unlike the corresponding potentials in Refs.~\cite{Cederbaum:2001,Johnson:2019qbe}, the Born term of this has the form of the usual folding potential, including contributions of three- and more-body forces.

As in single-channel approaches to the optical potential that take account of antisymmetry, the basis is not orthogonal. Various ways of dealing with this have been suggested by Feshbach \cite{Feshbach:1958nx,Feshbach:1962ut}, Kerman \cite{Kerman:1966}, and Capuzzi and Mahaux \cite{Capuzzi:1995}, all of which require constructing and then inverting the nontrivial matrix of overlaps of these states. In contrast, the version used here (which seems to be the only one not considered in Ref.~\cite{Capuzzi:1995}) does not require separate inversion of this matrix. Instead, it contains purely off-shell terms. These do not contribute directly to the scattering amplitude but, when the Lippmann-Schwinger equation is solved, they build up the factors that correct for the use of a nonorthogonal basis. 

The resulting optical potential has a, perhaps deceptively, simple form. Unlike the versions proposed by Feshbach \cite{Feshbach:1958nx,Feshbach:1962ut} or Capuzzi and Mahaux \cite{Capuzzi:1995}, its definition treats nonorthogonality in a symmetric way.  As a result the potential is Hermitian-analytic, which means that any non-Hermiticity for real energies arises only from open inelastic channels. It therefore has many of the properties of semi-phenomenological dispersive optical potentials \cite{Dickhoff:2018wdd}. However, there is still a price to be paid for the use of a nonorthogonal basis: the off-shell pieces of the potential depend linearly on energy. This growth with energy has led Capuzzi and Mahaux \cite{Capuzzi:1995} and Cederbaum \cite{Cederbaum:2001} to discard other versions with similar behaviour, and to concentrate instead on ones based on time-ordered Green's functions (see also \cite{Capuzzi:1996}). Nonetheless, the relatively simple structure of the resulting coupled equations suggests that it is worth exploring further as starting point for a Feshbach treatment, and hence as a possible source of optical potentials that could be used in reaction calculations.

The structure of the paper is as follows. In Sec.~2, I review how an optical potential can be constructed starting from the retarded Green's function. This is extended to coupled channels in Sec.~3. Then, in Sec.~4, I use a version of Feshbach's projection formalism to eliminate all but a small number of channels to generate potentials that can describe scattering and reactions in these channels.

\section{Retarded Green's function}

A scattering amplitude can be obtained from a Green's function by the usual 
Lehmann-Symanzik-Zimmermann (LSZ) reduction technique \cite{Lehmann:1954rq}. Here I start from the retarded Green's function, following the approach of Redish and Villars \cite{Redish:1970} and Johnson \cite{Johnson:2019qbe} for including recoil of the target nucleus. Compared with the work of Johnson, which was based on the corresponding wave functions, the Green's function yields a fully off-shell $T$ matrix and hence a more symmetric optical potential. This review of older approaches will establish the notation for the coupled-channel treatment below.

The retarded correlator of a creation and an annihilation operator between recoiling $A$-body states is
\begin{equation}
G_r({\bf k},{\bf k}';t-t')=-{\rm i}\,\theta(t-t'){\rm e}^{{\rm i}
(\epsilon({\bf k}')t'-\epsilon({\bf k})t)/A}\int\! {\rm d}^3{\bf K}\,
\langle{\bf K},0|a_{\bf k}(t)a^\dagger_{\bf k'}(t')
|{-}{\bf k}',0\rangle,
\end{equation}
where $|{-}{\bf k},0\rangle$ is the ground state of the nucleus (denoted by $0$), recoiling with momentum $-{\bf k}$. This state has energy $\epsilon({\bf k})/A$ relative to the energy $E_0$ of the $A$-body ground state. Note that all energies here are defined relative to $E_0$. The state created by acting the initial state with $a^\dagger_{\bf k'}$ has zero total momentum and so momentum conservation picks out ${\bf K}=-{\bf k}$. Spin quantum numbers on the nucleon operators are suppressed here, as are isospin ones in the coupled-channel extension below. Integrals over relative momenta should be understood as also containing sums over nucleon spins.

The time-dependent phase factor has been included in the definition of $G_r$ to make it a function of $t-t'$ only. It is needed because the initial and final $A$-body states can have different recoil energies in this off-shell correlator. A similar factor was introduced by Cederbaum \cite{Cederbaum:2001} and will be relevant to the coupled-channel case. Other choices of factor can be made but this one has the convenient feature of replacing the single-particle energies by the kinetic energies of relative motion, as in Ref.~\cite{Johnson:2019qbe}.

The operators here are in the Heisenberg representation, and they evolve as
\begin{equation}
a_{\bf k}(t)={\rm e}^{{\rm i}Ht}a_{\bf k}(0){\rm e}^{-{\rm i}Ht},
\end{equation}
with the full Hamiltonian 
\begin{equation}
H=H_0+V,
\end{equation}
where $H_0$ denotes the free single-particle Hamiltonian and the potential $V$ 
can includes two-, three- and more-body forces. 
Differentiating with respect to $t$ gives 
\begin{eqnarray}
&&\left({\rm i}\frac{\partial}{\partial t}-\frac{A+1}{A}\,\epsilon({\bf k})
\right)G_r({\bf k},{\bf k}';t-t')\\
&&\qquad =\delta(t-t'){\rm e}^{{\rm i}(\epsilon({\bf k}')t'-\epsilon({\bf k})t)/A}
\int\! {\rm d}^3{\bf K}\,
\langle{\bf K},0|a_{\bf k}(t)a^\dagger_{\bf k'}(t')
|{-}{\bf k}',0\rangle\\
&&\qquad\quad -\,{\rm i}\,\theta(t-t'){\rm e}^{{\rm i}(\epsilon({\bf k}')t'
-\epsilon({\bf k})t)/A}\int\! {\rm d}^3{\bf K}\,
\langle{\bf K},0|J_{\bf k}(t)a^\dagger_{\bf k'}(t')
|{-}{\bf k}',0\rangle,\nonumber
\end{eqnarray}
where $J_{\bf k}$ is the interaction ``current" defined as in Ref.~\cite{Villars:1967} by
\begin{equation}\label{eq:jident}
J_{\bf k}=\left[a_{\bf k},V\right].
\end{equation}

Differentiating again, and Fourier transforming with respect to $t-t'$, we get the LSZ reduction formula
\begin{eqnarray}\label{eq:lszgr}
&&\left(\omega-\frac{A+1}{A}\,\epsilon({\bf k}')\right)
\left(\omega-\frac{A+1}{A}\,\epsilon({\bf k})\right)
\widetilde G_r({\bf k},{\bf k}';\omega)\\
\noalign{\vspace{5pt}}
&&\qquad\quad = \left(\omega-\frac{A+1}{A}\,\epsilon({\bf k}')\right)
\int\! {\rm d}^3{\bf K}\,
\langle{\bf K},0|a_{\bf k}(0)a^\dagger_{\bf k'}(0)
|{-}{\bf k}',0\rangle
\cr
\noalign{\vspace{5pt}}
&&\qquad\qquad +\int\! {\rm d}^3{\bf K}
\langle{\bf K},0|J_{\bf k}(0)a^\dagger_{\bf k'}(0)
|{-}{\bf k}',0\rangle\cr
\noalign{\vspace{5pt}}
&&\qquad\qquad +\int\! {\rm d}^3{\bf K}\, 
\langle{\bf K},0|J_{\bf k}(0)\,
\frac{1}{\omega-H+{\rm i}\eta}\,J^\dagger_{\bf k'}(0)
|{-}{\bf k}',0\rangle,\nonumber
\end{eqnarray}
where $\eta$ is positive and infinitessimal.
After anticommuting the operators in the first two terms, this can be written
in the form
\begin{eqnarray}\label{eq:lsztr}
&&\left(\omega-\frac{A+1}{A}\,\epsilon({\bf k}')\right)
\left(\omega-\frac{A+1}{A}\,\epsilon({\bf k})\right)
\widetilde G_r({\bf k},{\bf k}';\omega)\\
\noalign{\vspace{5pt}}
&&\qquad\quad = \left(\omega-\frac{A+1}{A}\,\epsilon({\bf k})\right)
\delta^3({\bf k}-{\bf k}')+T_r({\bf k},{\bf k}';\omega),\nonumber
\end{eqnarray}
where the connected part, $T_r$, can be identified as the off-shell scattering
matrix. This is given by
\begin{eqnarray}\label{eq:tret}
T_r({\bf k},{\bf k}';\omega)
&=&\int\! {\rm d}^3{\bf K}\,
\langle{\bf K},0|V_{{\bf k}{\bf k}'}(0)
|{-}{\bf k}',0\rangle\\
\noalign{\vspace{5pt}}
&& +\int\! {\rm d}^3{\bf K}\, \langle{\bf K},0|J_{\bf k}(0)\,
\frac{1}{\omega-H+{\rm i}\eta}\,J^\dagger_{\bf k'}(0)
|{-}{\bf k}',0\rangle\cr
\noalign{\vspace{5pt}}
&& -\int\! {\rm d}^3{\bf K}\,
\langle{\bf K},0|a^\dagger_{\bf k'}(0)\,
\bigl(\omega-H-\epsilon({\bf k})-\epsilon({\bf k}')\bigr)
a_{\bf k}(0)|{-}{\bf k}',0\rangle,\nonumber
\end{eqnarray}
where 
\begin{equation}\label{eq:v1}
V_{{\bf k}{\bf k}'}=\bigl\{J_{\bf k},a^\dagger_{\bf k'}\bigr\}
=\bigl\{\left[a_{\bf k},V\right],a^\dagger_{\bf k'}\bigr\}.
\end{equation}
Here the identity \cite{Johnson:2019qbe}
\begin{equation}
J_{\bf k}(0)|{-}{\bf k}',0\rangle=\left(\frac{\epsilon({\bf k}')}{A}
-\epsilon({\bf k})-H\right)a_{\bf k}(0)|{-}{\bf k}',0\rangle
\end{equation}
has been used to put the final term of $T_r$ into a symmetric form.
Writing the potential $V$ as
\begin{eqnarray}
V&=&\frac{1}{4}\int {\rm d}^3{\bf k}_1\,{\rm d}^3{\bf k}_2\,{\rm d}^3{\bf k}'_1\,
{\rm d}^3{\bf k}'_2\,V_{2,\cal A}({\bf k}_1,{\bf k}_2;{\bf k}'_1,{\bf k}'_2)\,
a^\dagger_{{\bf k}_1}a^\dagger_{{\bf k}_2}a_{{\bf k}'_2}a_{{\bf k}'_1}\cr
\noalign{\vspace{5pt}}
&&+\frac{1}{36}\int {\rm d}^3{\bf k}_1\,{\rm d}^3{\bf k}_2\,{\rm d}^3{\bf k}_3\,
{\rm d}^3{\bf k}'_1\,{\rm d}^3{\bf k}'_2\,{\rm d}^3{\bf k}'_3\,
V_{3,\cal A}({\bf k}_1,{\bf k}_2,{\bf k}_3;{\bf k}'_1,{\bf k}'_2,{\bf k}'_3)\,
a^\dagger_{{\bf k}_1}a^\dagger_{{\bf k}_2}a^\dagger_{{\bf k}_3}
a_{{\bf k}'_3}a_{{\bf k}'_2}a_{{\bf k}'_1}+\cdots,
\end{eqnarray}
where $V_{n,\cal A}$ denotes the antisymmetrised matrix element of the $n$-body potential, the operator $V_{{\bf k}{\bf k}'}$ takes the form
\begin{eqnarray}\label{eq:v1b}
V_{{\bf k}{\bf k}'}&=&\int {\rm d}^3{\bf k}_2\,
{\rm d}^3{\bf k}'_2\,V_{2,\cal A}({\bf k},{\bf k}_2;{\bf k}',{\bf k}'_2)\,
a^\dagger_{{\bf k}_2}a_{{\bf k}'_2}\cr
\noalign{\vspace{5pt}}
&&+\frac{1}{4}\int {\rm d}^3{\bf k}_2\,{\rm d}^3{\bf k}_3\,
{\rm d}^3{\bf k}'_2\,{\rm d}^3{\bf k}'_3\,
V_{3,\cal A}({\bf k},{\bf k}_2,{\bf k}_3;{\bf k}',{\bf k}'_2,{\bf k}'_3)\,
a^\dagger_{{\bf k}_2}a^\dagger_{{\bf k}_3}
a_{{\bf k}'_3}a_{{\bf k}'_2}+\cdots,
\end{eqnarray}

The matrix element of the two-body piece of $V_{{\bf k}{\bf k}'}$ in the first term of Eq.~(\ref{eq:tret}) contains the direct or folding interaction of the projectile nucleon with the one of the target nucleons. This term also includes a contribution where the projectile interacts with two target nucleons via three-body forces. This generalises the usual folding potential to an integral over the two-body density. More generally, it will contain similar terms with three- and more-body densities, generated by many-body forces. Because of the antisymmetrisation of $V_{n,\cal A}$, all of these also include the corresponding exchange interactions.

The second term in Eq.~(\ref{eq:tret}) is the particle or dispersive contribution, where intermediate $(A+1)$-body states propagate. Finally, there is the hole term, also referred to as ``heavy-particle stripping" by Villars \cite{Villars:1967}. 
This term can also be expressed in terms of $J$s by using the identity (\ref{eq:jident}) to get
\begin{eqnarray}
T_r({\bf k},{\bf k}';\omega)
&=&\int\! {\rm d}^3{\bf K}\,\langle{\bf K},0|V_{{\bf k}{\bf k}'}(0)
|{-}{\bf k}',0\rangle\\
\noalign{\vspace{5pt}}
&& +\int\! {\rm d}^3{\bf K}\, \langle{\bf K},0|J_{\bf k}(0)\,
\frac{1}{\omega-H+{\rm i}\eta}\,J^\dagger_{\bf k'}(0)
|{-}{\bf k}',0\rangle\cr
\noalign{\vspace{5pt}}
&& -\int\! {\rm d}^3{\bf K}\,\langle{\bf K},0|J^\dagger_{\bf k'}(0)\,
\frac{\omega-H-\epsilon({\bf k})-\epsilon({\bf k}')}
{\bigl(\epsilon({\bf k})/A-\epsilon({\bf k}')-H\bigr)
\bigl(\epsilon({\bf k}')/A-\epsilon({\bf k})-H\bigr)}
J_{\bf k}(0)|{-}{\bf k}',0\rangle.\nonumber
\end{eqnarray}
This form makes it easy to see that, when $\omega$ is set equal to the initial 
energy, $\omega=(A+1)\,\epsilon({\bf k}')/A$, this agrees with the half-off-shell version in Eq.~(87) of Ref.~\cite{Johnson:2019qbe}.

A similar LSZ reduction applied to the time-ordered Green's function, as in 
Sec.~IV.B of Ref.~\cite{Johnson:2019qbe}, leads to the $T$ matrix
\begin{eqnarray}\label{eq:tto}
T({\bf k},{\bf k}';\omega)&=&
\int\! {\rm d}^3{\bf K}\,\langle{\bf K},0|V_{{\bf k}{\bf k}'}(0)
|-{\bf k}',0\rangle\cr
\noalign{\vspace{5pt}}
&& +\int\! {\rm d}^3{\bf K}\, \langle{\bf K},0|J_{\bf k}(0)\,
\frac{1}{\omega-H+{\rm i}\eta}\,J^\dagger_{\bf k'}(0)
|{-}{\bf k}',0\rangle\\
\noalign{\vspace{5pt}}
&& +\int\! {\rm d}^3{\bf K}\,\langle{\bf K},0|J^\dagger_{\bf k'}(0)\,
\frac{1}{\omega+H-(\epsilon({\bf k})+\epsilon({\bf k}'))/A-{\rm i}\eta}\,
J_{\bf k}(0)|{-}{\bf k}',0\rangle.\nonumber
\end{eqnarray}
The two versions agree on-shell, where $\omega=(A+1)\,\epsilon({\bf k})/A$ 
and $\epsilon({\bf k}')=\epsilon({\bf k})$. 

The $(A+1)$-body propagator in the particle term includes the channel with one nucleon in the presence of the $A$-nucleon ground state. It is therefore reducible with respect to this channel. An irreducible self-energy, which may 
be interpreted as an optical potential, can be obtained by writing $T_r$ as the solution to a Lippmann-Schwinger equation in the effective one-body space of states of relative motion:
\begin{equation}\label{eq:lstret}
T_r=\Sigma_r+\Sigma_r\,G_0\,T_r,
\end{equation}
where the free propagator is
\begin{equation}
G_0({\bf k},{\bf k}';\omega)=\frac{\delta^3({\bf k}-{\bf k}')}{\displaystyle\omega-\frac{A+1}{A}\,
\epsilon({\bf k})+{\rm i}\eta}.
\end{equation}
From Eq.~(\ref{eq:lsztr}), this can be seen to be equivalent to the definition from the inverse of the Green's function,
\begin{equation}\Sigma_r=G_0^{-1}-G_r^{-1}.
\end{equation}

Capuzzi and Mahaux point out that a possible issue with defining an optical potential in this way is that the retarded Green's function can be singular if there any completely occupied orbitals \cite{Capuzzi:1995} (see also Ref.~\cite{Cederbaum:2001}). In that case, the Lippmann-Schwinger equation (\ref{eq:lstret}) cannot be inverted to obtain an optical potential. Here, the problem with singularity of operators such as $\widetilde G_r$ does not arise since including centre-of-mass motion ensures that the $A$-body state is not a pure Slater determinant.  

The self-energy defined in this way is Hermitian-analytic \cite{Capuzzi:1995}, like dispersive optical potentials. However, unlike those potentials, $\Sigma_r$ grows linearly with the off-shell energy $\omega$, as can be seen from the final term in the retarded $T$ matrix (\ref{eq:tret}). This is similar to the potential in Eq.~(4.33) of Ref.~\cite{Capuzzi:1995}. Although the feature is strongly deprecated by both Capuzzi and Mahaux \cite{Capuzzi:1995} and Cederbaum \cite{Cederbaum:2001}, it is just part of the price for working with the nonorthogonal basis, as discussed further in the next section.

Although the form of the Born terms in $T_r$ implies that the corresponding optical potential does not vanish in the limit of noninteracting particles, the pieces responsible are purely off-shell in nature and they do not give rise to scattering in the absence of interactions. This can be seen from the Green's function: switching off the interaction currents in Eq.~(\ref{eq:lszgr}) leads to
\begin{equation}
\widetilde G_r({\bf k},{\bf k}';\omega)=\frac{1}{\displaystyle\omega
-\frac{A+1}{A}\,\epsilon({\bf k})+{\rm i}\eta}\,\int\! {\rm d}^3{\bf K}\,
\langle{\bf K},0|a_{\bf k}(0)a^\dagger_{\bf k'}(0)
|{-}{\bf k}',0\rangle.
\end{equation}
This is just free propagation, taking account of the nonorthogonal nature of the basis states. Similar off-shell behaviour will play a role in the treatment below.

A more pragmatic concern was raised by Johnson at the end of 
Ref.~\cite{Johnson:2019qbe}, namely that, if one has the full $T$ matrix (or Green's function) then the distorted waves can be constructed from it directly, without the need to convert it into an optical potential. This motivates looking for an extension of this approach that could lead to tools for constructing potentials that could make contact with theories of direct reactions. The coupled-channel version in the next section offers a possibility of doing this, at least for inelastic scattering and charge-exchange reactions.

\section{Coupled channels}

The approach of the previous section can be extended to matrix elements of creation and annihilation operators between general states of the $A$-nucleon system, as in Ref.~\cite{Johnson:2019qbe}. A similar matrix of 
off-diagonal or inelastic Green's functions was previously studied by
Cederbaum \cite{Cederbaum:2001}, but without considering recoil. The recoiling $A$-body states are denoted here by $|{-}{\bf k}\,n\rangle$ and have momenta $-{\bf k}$ and energies $\Delta_n+\epsilon({\bf k})/A$. These define a space that I refer to as the ``effective one-body space", consisting of channels 
each of which describes the relative motion of the projectile and a particular state of the target.\footnote{This is the ``extended $B$-space" of Ref.~\cite{Johnson:2019qbe}.} The matrix of retarded Green's functions is 
\begin{eqnarray}
&&{\cal G}_r({\bf k}\,n,{\bf k}'\,n';t-t')\\
\noalign{\vspace{5pt}}
&&\qquad=-{\rm i}\,\theta(t-t'){\rm e}^{{\rm i}
[(\epsilon({\bf k}')/A+\Delta_n')t'-(\epsilon({\bf k})/A+\Delta_n)t)]}
\int\! {\rm d}^3{\bf K}\,
\langle{\bf K}\,n|a_{\bf k}(t)a^\dagger_{\bf k'}(t')
|{-}{\bf k}'\,n'\rangle.
\nonumber
\end{eqnarray}
In this case, isospin as well as spin labels are suppressed but the creation and annihilation operators should be understood to carry them, allowing the approach to treat charge exchange as well as inelastic scattering. The time-dependent phase factors again make these functions of $t-t'$ only, as in Ref.~\cite{Cederbaum:2001}.

Taking the $t$ derivative of ${\cal G}_r$ and rearranging terms gives
\begin{eqnarray}
&&\left({\rm i}\frac{\partial}{\partial t}-\frac{A+1}{A}\,\epsilon({\bf k})
-\Delta_n\right){\cal G}_r({\bf k}\,n,{\bf k}'\,n';t-t')\\
\noalign{\vspace{5pt}}
&&\qquad\quad =\delta(t-t'){\rm e}^{{\rm i}
[\epsilon({\bf k}')/A+\Delta_n'-\epsilon({\bf k})/A-\Delta_n]t}
\int\! {\rm d}^3{\bf K}\,
\langle{\bf K}\,n|a_{\bf k}(t)a^\dagger_{\bf k'}(t)
|{-}{\bf k}'\,n'\rangle\cr
\noalign{\vspace{5pt}}
&&\qquad\qquad -\,{\rm i}\,\theta(t-t'){\rm e}^{{\rm i}
[(\epsilon({\bf k}')/A+\Delta_n')t'-(\epsilon({\bf k})/A+\Delta_n)t)]}
\int\! {\rm d}^3{\bf K}\,
\langle{\bf K}\,n|J_{\bf k}(t)a^\dagger_{\bf k'}(t')
|{-}{\bf k}'\,n'\rangle.\nonumber
\end{eqnarray}
As above, a second time derivative could be used to extract the corresponding scattering matrix. Alternatively a set of coupled equations for the Green's functions can be obtained by Fourier transforming to get
\begin{eqnarray}\label{eq:eqgrt}
&&\left(\omega-\frac{A+1}{A}\,\epsilon({\bf k})
-\Delta_n\right)\widetilde {\cal G}_r({\bf k}\,n,{\bf k}'\,n';\omega)\\
\noalign{\vspace{5pt}}
&&\qquad\quad =\int\! {\rm d}^3{\bf K}\,
\langle{\bf K}\,n|a_{\bf k}(0)a^\dagger_{\bf k'}(0)
|{-}{\bf k}'\,n'\rangle\cr
\noalign{\vspace{5pt}}
&&\qquad\qquad +\int\! {\rm d}^3{\bf K}\, 
\langle{\bf K}\,n|J_{\bf k}(0)\,
\frac{1}{\omega-H+{\rm i}\eta}\,a^\dagger_{\bf k'}(0)
|{-}{\bf k}'\,n'\rangle.\nonumber
\end{eqnarray}

At this point, both Cederbaum \cite{Cederbaum:2001} and Johnson \cite{Johnson:2019qbe} insert a complete set of of $A$-body states between the annihilation operators inside $J_{\bf k}$. The resulting potential-like operator has the form
\begin{eqnarray}
V'_{{\bf k}{\bf k}'}&=&\frac{1}{2}\int {\rm d}^3{\bf k}_2\,
{\rm d}^3{\bf k}'_2\,V_{2,\cal A}({\bf k},{\bf k}_2;{\bf k}',{\bf k}'_2)\,
a^\dagger_{{\bf k}_2}a_{{\bf k}'_2}\cr
\noalign{\vspace{5pt}}
&&+\frac{1}{12}\int {\rm d}^3{\bf k}_2\,{\rm d}^3{\bf k}_3\,
{\rm d}^3{\bf k}'_2\,{\rm d}^3{\bf k}'_3\,
V_{3,\cal A}({\bf k},{\bf k}_2,{\bf k}_3;{\bf k}',{\bf k}'_2,{\bf k}'_3)\,
a^\dagger_{{\bf k}_2}a^\dagger_{{\bf k}_3}
a_{{\bf k}'_3}a_{{\bf k}'_2}+\cdots.
\end{eqnarray}
This differs from the one in Eq.~(\ref{eq:v1b}) by having different combinatoric factors: its two-body term is multiplied by $1/2$, and the three-body by $1/12$ instead of $1/4$. (Compare Eq.~(3.12) of Ref.~\cite{Cederbaum:2001} or (59) of Ref.~\cite{Johnson:2019qbe} for the same factor of 1/2 in the two-body term.) As a result, this approach has no simple connection to a folding potential, generalised to include three- or more-body forces.

Instead, I introduce a complete set of the $(A+1)$-body states, $a^\dagger_{\bf k}(0)|{-}{\bf k}\,n\rangle$. The states are not orthogonal but have an overlap matrix
\begin{equation}
\int\! {\rm d}^3{\bf K}\, 
\langle{\bf K}\,n|a_{\bf k}(0)a^\dagger_{\bf k'}(0)|{-}{\bf k}'\,n'\rangle
={\cal I}({\bf k}\,n,{\bf k}'\,n')-{\cal K}({\bf k}\,n,{\bf k}'\,n'),
\end{equation}
where
\begin{equation}
{\cal I}({\bf k}\,n,{\bf k}'\,n')=\delta_{nn'}\delta({\bf k}-{\bf k}')
\end{equation}
is the identity operator in the effective one-body space, and 
\begin{equation}
{\cal K}({\bf k}\,n,{\bf k}'\,n')=\int\! {\rm d}^3{\bf K}\, 
\langle{\bf K}\,n|a^\dagger_{\bf k'}(0)a_{\bf k}(0)|{-}{\bf k}'\,n'\rangle.
\end{equation}
This overlap matrix is the multi-channel version of the operator $1-K$ introduced by Feshbach \cite{Feshbach:1958nx,Feshbach:1962ut} (see also Capuzzi and Mahaux \cite{Capuzzi:1995}). In terms of this, the identity operator in the space of states with zero total momentum has the form
\begin{equation}
I=\sum_{n,n'}\int\! {\rm d}^3{\bf k}\,{\rm d}^3{\bf k}'\,{\rm d}^3{\bf K}\;
a^\dagger_{\bf k'}(0)|{-}{\bf k}'\,n'\rangle\,
({\cal I}-{\cal K})^{-1}({\bf k}'\,n',{\bf k}\,n)\,
\langle{\bf K}\,n|a_{\bf k}(0),
\end{equation}
where the inverse is defined in the effective one-body space. As noted above, questions about singularity of the operator are avoided by the inclusion of recoil which means that the $A$-body states are not pure Slater determinants.

Inserting this identity to the left of the $(A+1)$-body Green's function in the final term of Eq.~(\ref{eq:eqgrt}) gives a set of coupled equations for $\widetilde {\cal G}_r$,
\begin{eqnarray}
&&\left(\omega-\frac{A+1}{A}\,\epsilon({\bf k})
-\Delta_n\right)\widetilde {\cal G}_r({\bf k}\,n,{\bf k}'\,n';\omega)\\
\noalign{\vspace{5pt}}
&&\qquad\quad ={\cal I}({\bf k}\,n,{\bf k}'\,n')
-{\cal K}({\bf k}\,n,{\bf k}'\,n')\cr
\noalign{\vspace{5pt}}
&&\qquad\qquad+\sum_{n'',n'''}\int\!{\rm d}^3{\bf k}''\,{\rm d}^3{\bf k}'''\,
{\rm d}^3{\bf K}\,{\rm d}^3{\bf K}'''\; 
\langle{\bf K}\,n|J_{\bf k}(0)a^\dagger_{\bf k''}(0)|{-}{\bf k}''\,n''\rangle\cr
\noalign{\vspace{5pt}}
&&\hspace{5.5cm}
\times\,({\cal I}-{\cal K})^{-1}({\bf k}''\,n'',{\bf k}'''\,n''')\,
\widetilde {\cal G}_r({\bf K'''}\,n''',{\bf k}'\,n';\omega).\nonumber
\end{eqnarray}
More schematically, this can be written entirely in terms of operators in the effective one-body space as
\begin{equation}\label{eq:ccegr}
\bigl(\omega{\cal I}-{\cal H}_0\bigr)\widetilde {\cal G}_r(\omega)
={\cal I}-{\cal K}+{\cal U}\,({\cal I}-{\cal K})^{-1}
\widetilde {\cal G}_r(\omega),
\end{equation}
where ${\cal H}_0$ is the diagonal matrix of effective one-body energies,
\begin{equation}
{\cal H}_0({\bf k}\,n,{\bf k}'\,n')=\left(\frac{A+1}{A}\,\epsilon({\bf k})
+\Delta_n\right)\delta_{nn'}\delta({\bf k}-{\bf k}'),
\end{equation}
and the potential ${\cal U}$ has matrix elements
\begin{equation}
{\cal U}({\bf k}\,n,{\bf k}'\,n')=\int\! {\rm d}^3{\bf K}\, 
\langle{\bf K}\,n|J_{\bf k}(0)a^\dagger_{\bf k'}(0)|{-}{\bf k}'\,n'\rangle.
\end{equation}

Multiplying Eq.~(\ref{eq:ccegr}) from the right by $({\cal I}-{\cal K})^{-1}$, the set of equations for $\widetilde {\cal G}_r$ can be put into the more symmetric form
\begin{equation}\label{eq:ccsym}
\bigl(\omega{\cal I}-{\cal H}_0-{\cal V}\bigr)({\cal I}-{\cal K})^{-1}
\widetilde {\cal G}_r(\omega)({\cal I}-{\cal K})^{-1}={\cal I},
\end{equation}
where the potential ${\cal V}$ is given by
\begin{equation}\label{eq:vu}
{\cal V}=\bigl(\omega{\cal I}-{\cal H}_0\bigr){\cal K}+{\cal U}.
\end{equation}
After anticommuting the operators inside ${\cal U}$, this potential becomes
\begin{eqnarray}
{\cal V}({\bf k}\,n,{\bf k}'\,n')&=&
\left(\omega-\frac{A+1}{A}\,\epsilon({\bf k})-\Delta_n\right)
\int\! {\rm d}^3{\bf K}\, 
\langle{\bf K}\,n|a^\dagger_{\bf k'}(0)a_{\bf k}(0)|{-}{\bf k}'\,n'\rangle\\
\noalign{\vspace{5pt}}
&&-\int\! {\rm d}^3{\bf K}\, 
\langle{\bf K}\,n|a^\dagger_{\bf k'}(0)J_{\bf k}(0)|{-}{\bf k}'\,n'\rangle
+\int\! {\rm d}^3{\bf K}\, 
\langle{\bf K}\,n|V_{{\bf k}{\bf k}'}(0)|{-}{\bf k}'\,n'\rangle.\nonumber
\end{eqnarray}
Then, with the aid of the generalised version of the identity (\ref{eq:jident}),
\begin{equation}
J_{\bf k}(0)|{-}{\bf k}',n'\rangle=\left(\frac{\epsilon({\bf k}')}{A}+\Delta_{n'}
-\epsilon({\bf k})-H\right)a_{\bf k}(0)|{-}{\bf k}',n'\rangle,
\end{equation}
it can be written in the form
\begin{eqnarray}
{\cal V}({\bf k}\,n,{\bf k}'\,n')
&=&\int\! {\rm d}^3{\bf K}\, 
\langle{\bf K}\,n|V_{{\bf k}{\bf k}'}(0)|{-}{\bf k}'\,n'\rangle\\
\noalign{\vspace{5pt}}
&&+\int\! {\rm d}^3{\bf K}\, 
\langle{\bf K}\,n|a^\dagger_{\bf k'}(0)\bigl(\omega+H-\Delta_n-\Delta_{n'}
-\epsilon({\bf k})/A
- \epsilon({\bf k}')/A\bigr)a_{\bf k}(0)|{-}{\bf k}'\,n'\rangle,\nonumber
\end{eqnarray}
showing that it is Hermitian-analytic, ${\cal V}(\omega)^\dagger={\cal V}(\omega^*)$.

The terms in the potential ${\cal V}$ can be seen to be multichannel versions of the Born terms in the retarded $T$ matrix, Eq.~(\ref{eq:tret}): a ``folding potential" (including exchange), and a ``heavy-particle stripping" term. The latter can be expressed in terms of interaction currents as 
\begin{eqnarray}\label{eq:vj}
{\cal V}({\bf k}\,n,{\bf k}'\,n')
&=&\int\! {\rm d}^3{\bf K}\, 
\langle{\bf K}\,n|V_{{\bf k}{\bf k}'}(0)|{-}{\bf k}'\,n'\rangle\\
\noalign{\vspace{5pt}}
&&+\int\! {\rm d}^3{\bf K}\, 
\langle{\bf K}\,n|J^\dagger_{\bf k'}(0)\frac{\omega+H-\Delta_n-\Delta_{n'}
-\epsilon({\bf k})/A- \epsilon({\bf k}')/A}
{\bigl(\epsilon({\bf k})/A-\Delta_n-\epsilon({\bf k}')-H\bigr)
\bigl(\epsilon({\bf k}')/A-\Delta_{n'}-\epsilon({\bf k})-H\bigr)}
J_{\bf k}(0)|{-}{\bf k}'\,n'\rangle\nonumber
\end{eqnarray}
Alternative forms of the effective one-body equation that could also be used to define an optical potential are discussed in Ref.~\cite{Capuzzi:1995}. They are summarised in the Appendix, expressed in the coupled-channel formalism used here.

The potential ${\cal V}$ provides the starting point for construction of an optical potential in the next section, by eliminating all but one or a small number of channels. If, instead, one were to simply restrict the effective one-body space to
a subspace containing only a few $A$-body states, then the equations (\ref{eq:ccsym}) would reduce to a version of the resonating group method (RGM) \cite{Wheeler:1937zz,Saito:1977,Kamimura:1977,Tang:1978zz}. 

\section{Projection, scattering and reactions}

The matrix $\widetilde{\cal G}_r$ of Green's functions can be written in the form
\begin{equation}
\widetilde{\cal G}_r=({\cal I}-{\cal K})\widehat{\cal G}_r({\cal I}-{\cal K}),
\end{equation}
where 
\begin{equation}
\widehat{\cal G}_r=\frac{1}{\omega{\cal I}-{\cal H}_0-{\cal V}+{\rm i}\eta}\,.
\end{equation}
This modified Green's function satisfies a simpler set of coupled-channel Lippmann-Schwinger (LS) equations without explicit overlap matrices,
\begin{equation}
\widehat{\cal G}_r=\widehat{\cal G}_0
+\widehat{\cal G}_0\,{\cal V}\,\widehat{\cal G}_r,
\end{equation} 
where the propagator for free relative motion is
\begin{equation}
\widehat{\cal G}_0=\frac{1}{\omega{\cal I}-{\cal H}_0+{\rm i}\eta}. 
\end{equation} 
With the aid of this and the definition (\ref{eq:vu}) of ${\cal V}$, the left-hand factor of ${\cal I}-{\cal K}$ in $\widetilde{\cal G}_r$ can be absorbed, giving 
\begin{equation}
\widetilde{\cal G}_r=\Bigl(\widehat{\cal G}_0
+\widehat{\cal G}_0\,{\cal U}\,\widehat{\cal G}_r\Bigr)
({\cal I}-{\cal K}).
\end{equation}
Repeating the process with the LS equations with their kernel on the right absorbs the second factor and leaves
\begin{equation}\label{eq:grnok}
\widetilde{\cal G}_r=\widehat{\cal G}_0
+\widehat{\cal G}_0\Bigl[\widehat{\cal V}
+{\cal U}\,\widetilde{\cal G}_r\,\overline{\cal U}
\Bigr]\widehat{\cal G}_0,
\end{equation}
where the modified potentials are
\begin{eqnarray}
\widehat{\cal V}&=&{\cal U}-{\cal K}(\omega{\cal I}-{\cal H}_0)\\
\noalign{\vspace{5pt}}
&=&{\cal V}-(\omega{\cal I}-{\cal H}_0){\cal K}-{\cal K}
(\omega{\cal I}-{\cal H}_0),\\
\noalign{\vspace{5pt}}
\overline{\cal U}&=&{\cal V}-{\cal K}(\omega{\cal I}-{\cal H}_0).
\end{eqnarray}

The factor inside the square bracket in Eq.~(\ref{eq:grnok}) is the scattering matrix in the effective one-body space:
\begin{equation}
\widehat{\cal T}=\widehat{\cal V}+{\cal U}\,\widetilde{\cal G}_r\,
\overline{\cal U}.
\end{equation}
The potentials $\widehat{\cal V}$, ${\cal U}$ and $\overline{\cal U}$ differ 
from ${\cal V}$ only by terms like ${\cal K}(\omega{\cal I}-{\cal H}_0)$. These vanish when they act on on-shell effective one-body states and so they affect only the off-shell behaviour of the $T$ matrix.
The same on-shell scattering amplitudes can therefore be obtained from the alternative $T$ matrix defined using the Hermitian-analytic potential ${\cal V}$ everywhere,
\begin{equation}
{\cal T}={\cal V}+{\cal V}\,\widetilde{\cal G}_r\,{\cal V}.
\end{equation}
It is worth noting that, in contrast to other approaches \cite{Feshbach:1958nx,Feshbach:1962ut,Capuzzi:1995,Kerman:1966}, there is no need to invert the overlap matrix ${\cal I}-{\cal K}$ in order to construct the potential ${\cal V}$. In the version here, this inversion is done implicitly when the LS equation is solved for $\widetilde{\cal G}_r$ or ${\cal T}$. 

Having defined a $T$ matrix entirely in terms of operators in the effective
one-body space, we can apply Feshbach's approach \cite{Feshbach:1958nx,Feshbach:1962ut} in this space. This introduces a projector ${\cal P}$ onto a subspace, usually called the ``model space". For example, if this subspace consists purely of the channel with the $A$-body target in ground state, it yields an optical potential for elastic scattering. The projector onto the complementary subspace is
\begin{equation}
{\cal Q}={\cal I}-{\cal P}.
\end{equation}
In the usual way, the projected $T$ matrix can be shown to satisfy the LS equation
\begin{equation}\label{eq:lstp}
{\cal P}{\cal T}{\cal P}={\cal V}_{\cal P}+{\cal V}_{\cal P}\,
\widehat{\cal G}_0\,{\cal P}{\cal T}{\cal P},
\end{equation}
where the optical potential in the ${\cal P}$-space is, in terms of the 
potential ${\cal V}$,
\begin{equation}\label{eq:vop}
{\cal V}_{\cal P}={\cal P}{\cal V}{\cal P}+{\cal P}{\cal V}{\cal Q}\,
\frac{1}{\omega{\cal I}-{\cal H}_0-{\cal Q}{\cal V}{\cal Q}+{\rm i}\eta}\,
{\cal Q}{\cal V}{\cal P}.
\end{equation}
This result, together with the expression (\ref{eq:vj}) for the potential ${\cal V}$, forms the main result of the approach developed here. This optical potential is Hermitian-analytic but it inherits the linear dependence on energy from ${\cal V}$.

The projected $T$ matrix can describe not just elastic scattering but also reactions such as inelastic scattering or charge exchange. To illustrate this, consider projection onto the channels where the $A$-body nucleus is in one of two states. One of these is the ground state of the $A$-body target, while the other could be either an excited state of the target or a state of the nucleus with a neutron swapped for a proton. The projection operator can be written
\begin{equation}
{\cal P}={\cal P}_0+{\cal P}_1,
\end{equation}
where ${\cal P}_0$ projects onto the elastic channel and ${\cal P}_1$ onto the reaction channel. It is also useful to define the projector onto all reaction channels,
\begin{equation}
{\cal Q}_0={\cal I}-{\cal P}_0.
\end{equation}

From Eq.~(\ref{eq:lstp}), the reaction amplitude ${\cal P}_1{\cal T}{\cal P}_0$
satisfies the LS equation
\begin{equation}
{\cal P}_1{\cal T}{\cal P}_0={\cal P}_1{\cal V}_{\cal P}{\cal P}_0
+{\cal P}_1{\cal V}_{\cal P}{\cal P}_0\,\widehat{\cal G}_0\,
{\cal P}_0{\cal T}{\cal P}_0
+{\cal P}_1{\cal V}_{\cal P}{\cal P}_1\,\widehat{\cal G}_0\,
{\cal P}_1{\cal T}{\cal P}_0.
\end{equation}
The solution to this can be written in the form
\begin{equation}\label{eq:t10}
{\cal P}_1{\cal T}{\cal P}_0=\Bigl(1+{\cal P}_1{\cal T}_{1}{\cal P}_1\,
\widehat{\cal G}_0\Bigr){\cal P}_1{\cal V}_{\cal P}{\cal P}_0
\Bigl(1+\widehat{\cal G}_0\,{\cal P}_0{\cal T}{\cal P}_0\Bigr),
\end{equation}
where ${\cal T}_{1}$ satisfies the LS equation
\begin{equation}
{\cal P}_1{\cal T}_{1}{\cal P}_1={\cal P}_1{\cal V}_{\cal P}{\cal P}_1
+{\cal P}_1{\cal V}_{\cal P}{\cal P}_1\,\widehat{\cal G}_0\,
{\cal P}_1{\cal T}_1
{\cal P}_1.
\end{equation}
This amplitude has a familiar distorted-wave structure. It is expressed here in a ``post" form, where the reaction is taken to occur at the last point the system scatters out of the elastic channel. The factor on the right contains the full scattering in the elastic channel. On-shell this gives the incoming distorted wave in that channel. This could be obtained from an optical potential with all reaction channels eliminated, which has a similar form to the one in Eq.~(\ref{eq:vop}), but with ${\cal Q}$ replaced by ${\cal Q}_0$. 
The factor on the left generates a distorted wave but with a modified optical potential, ${\cal P}_1{\cal V}_{\cal P}{\cal P}_1$, where all but two channels (the elastic and relevant reaction) have been eliminated. The transition potential ${\cal P}_1{\cal V}_{\cal P}{\cal P}_1$ also includes the effects of intermediate channels outside the two-channel model space.

\section{Summary}

This work has described a possible approach to constructing an optical potential that includes the effects of antisymmetry and target recoil. Many approaches to this problem have been suggested over the years. The present one shares features with some of them, but also has distinctive differences. Unlike, for example Feshbach's original approach \cite{Feshbach:1958nx}, it treats nonorthogonality symmetrically and so the potential is Hermitian-analytic, like phenomological ones. It differs from related potentials proposed by Cederbaum \cite{Cederbaum:2001} and, more recently Johnson \cite{Johnson:2019qbe}, by retaining a simple connection to folding potentials, even when three or more-body forces are present.

Microscopic approaches to optical potentials commonly start from the time-ordered Green's function \cite{Capuzzi:1995,Capuzzi:1996,Cederbaum:2001,Dickhoff:2018wdd}, which means they can be linked to many-body calculations of target structure. In contrast, the version here is obtained from the retarded Green's function. It also includes recoil of the target nucleus \cite{Redish:1970,Johnson:2017xdg,Johnson:2019qbe}. Both of the features could make it a more natural framework for describing direct reactions on light nuclei.

However the nonorthogonality of the basis does lead to a potential that has a
linear dependence on energy. Although this feature has led Mahaux and others to discard such potentials and to focus on versions based on the time-ordered Green's function \cite{Capuzzi:1995,Cederbaum:2001,Capuzzi:1996}, the structures of the these off-shell terms are similar to ones that appear in the RGM \cite{Wheeler:1937zz,Saito:1977,Kamimura:1977,Tang:1978zz}. The way they appear also means that there is no need to separately invert the overlap matrix in order to construct the potential.

By applying a version of Feshbach's projection formalism \cite{Feshbach:1958nx,Feshbach:1962ut} to the couped-channel version, all but a small number of channels can be eliminated. This generates optical potentials that can describe scattering and reactions in the channels of interest.

Despite these encouraging features, further work still needs to be done to implement this as a practical approach. In particular, it needs to be extended to cover rearrangement reactions, which will introduce issues of overcompleteness as well as nonorthogonality of the basis \cite{Cotanch:1976zz,Birse:1982ih}.

\section*{Acknowledgements}

I am grateful to J. Kirscher, R. Johnson and N. Timofeyuk for the conversations that prompted this work, and to R. Johnson for further helpful discussions about the ideas involved. 
This work was supported by the UK STFC under grant ST/P004423/1.
\appendix

\section{Alternative definitions of the optical potential}

Alternative versions of the effective one-body equation, which handle nonothogonality in different ways, can be found in Ref.~\cite{Capuzzi:1995} for the single-channel case. Any of them can be used as the starting point for a  Feshbach treatment to produce an optical potential. Their coupled-channel versions are given here, in forms that allow them to be related to the one in Sec.~III.

The potential ${\cal V}$ defined above was obtained from the equation (\ref{eq:ccegr}) for the retarded Green's function by multiplying it from the right by $({\cal I}-{\cal K})^{-1}$. This leads to a Hermitian-analytic optical potential. An alternative is to multiply from the left by $({\cal I}-{\cal K})^{-1}$. This puts the equation for $\widetilde{\cal G}_r$ into the, also symmetric, form
\begin{equation}
({\cal I}-{\cal K})^{-1}\Bigl[\bigl(\omega{\cal I}-{\cal H}_0\bigr)
({\cal I}-{\cal K})-{\cal U}\Bigr]
({\cal I}-{\cal K})^{-1}\,\widetilde{\cal G}_r={\cal I},
\end{equation}
which is the coupled-channel equivalent of Eq.~(4.32f) of Ref.~\cite{Capuzzi:1995}.\footnote{Note that there is an error there: the right-hand side of (4.32c) should be $1-K$ and hence the factors in (4.32f) should be $(1-K)^{-1}$. Despite this, their Eq.~(4.33) is correct.} This can be rewritten as
\begin{equation}
\Bigl[\omega{\cal I}-{\cal H}_0-{\cal V}_{CM}\Bigr]
\widetilde{\cal G}_r={\cal I}.
\end{equation}
where the potential that corresponds to the first two terms of their Eq.~(4.33) is
\begin{equation}
{\cal V}_{CM}=-({\cal I}-{\cal K})^{-1}{\cal K}
\bigl(\omega{\cal I}-{\cal H}_0\bigr)+({\cal I}-{\cal K})^{-1}
{\cal U}({\cal I}-{\cal K})^{-1}.
\end{equation}
Like ${\cal V}$ defined above, this contains a linear dependence on the energy $\omega$, as discussed in Ref.~\cite{Capuzzi:1995}.

Feshbach's original approach avoids this linear dependence on energy \cite{Feshbach:1958nx,Feshbach:1962ut}. It simply absorbs the factor of $({\cal I}-{\cal K})^{-1}$ in Eq.~(\ref{eq:ccegr}) into the potential:
\begin{equation}
{\cal V}_F={\cal U}({\cal I}-{\cal K})^{-1}.
\end{equation}
The asymmetric definition means that this is not Hermitian-analytic and neither is the resulting optical potential.

Finally, Kerman proposed a definition that is Hermitian-analytic and avoids a linear dependence on energy \cite{Kerman:1966}. This multiplies Eq.~(\ref{eq:ccegr}) by factors of the square root of $({\cal I}-{\cal K})^{-1/2}$, putting it into the form
\begin{equation}
\bigl(\omega{\cal I}-{\cal H}_0-{\cal V}_K\bigr)({\cal I}-{\cal K})^{-1/2}\,
\widetilde{\cal G}_r\,({\cal I}-{\cal K})^{-1/2}={\cal I},
\end{equation}
where
\begin{equation}
{\cal V}_K=({\cal I}-{\cal K})^{-1/2}\,{\cal H}_0\,({\cal I}-{\cal K})^{1/2}-{\cal H}_0
+({\cal I}-{\cal K})^{-1/2}\,{\cal U}\,({\cal I}-{\cal K})^{-1/2}.
\end{equation}

\end{document}